\documentclass[aps,prl,twocolumn,showpacs,superscriptaddress,groupedaddress]{revtex4}  
\usepackage{graphicx}  
\usepackage{dcolumn}   
\usepackage{bm}        
\usepackage{amssymb}   
\usepackage{amsmath}

\def\d0{D\O}

\begin{document}



\hspace{5.2in} \mbox{FERMILAB-PUB-05/PUB-11-563-E}

\title{Measurement of the relative branching ratio
of \boldmath $B^0_s \rightarrow J/\psi f_{0}(980)$ to $B_{s}^{0} \rightarrow J/\psi \phi$}

%
\affiliation{Universidad de Buenos Aires, Buenos Aires, Argentina}
\affiliation{LAFEX, Centro Brasileiro de Pesquisas F{\'\i}sicas, Rio de Janeiro, Brazil}
\affiliation{Universidade do Estado do Rio de Janeiro, Rio de Janeiro, Brazil}
\affiliation{Universidade Federal do ABC, Santo Andr\'e, Brazil}
\affiliation{Instituto de F\'{\i}sica Te\'orica, Universidade Estadual Paulista, S\~ao Paulo, Brazil}
\affiliation{University of Science and Technology of China, Hefei, People's Republic of China}
\affiliation{Universidad de los Andes, Bogot\'{a}, Colombia}
\affiliation{Charles University, Faculty of Mathematics and Physics, Center for Particle Physics, Prague, Czech Republic}
\affiliation{Czech Technical University in Prague, Prague, Czech Republic}
\affiliation{Center for Particle Physics, Institute of Physics, Academy of Sciences of the Czech Republic, Prague, Czech Republic}
\affiliation{Universidad San Francisco de Quito, Quito, Ecuador}
\affiliation{LPC, Universit\'e Blaise Pascal, CNRS/IN2P3, Clermont, France}
\affiliation{LPSC, Universit\'e Joseph Fourier Grenoble 1, CNRS/IN2P3, Institut National Polytechnique de Grenoble, Grenoble, France}
\affiliation{CPPM, Aix-Marseille Universit\'e, CNRS/IN2P3, Marseille, France}
\affiliation{LAL, Universit\'e Paris-Sud, CNRS/IN2P3, Orsay, France}
\affiliation{LPNHE, Universit\'es Paris VI and VII, CNRS/IN2P3, Paris, France}
\affiliation{CEA, Irfu, SPP, Saclay, France}
\affiliation{IPHC, Universit\'e de Strasbourg, CNRS/IN2P3, Strasbourg, France}
\affiliation{IPNL, Universit\'e Lyon 1, CNRS/IN2P3, Villeurbanne, France and Universit\'e de Lyon, Lyon, France}
\affiliation{III. Physikalisches Institut A, RWTH Aachen University, Aachen, Germany}
\affiliation{Physikalisches Institut, Universit{\"a}t Freiburg, Freiburg, Germany}
\affiliation{II. Physikalisches Institut, Georg-August-Universit{\"a}t G\"ottingen, G\"ottingen, Germany}
\affiliation{Institut f{\"u}r Physik, Universit{\"a}t Mainz, Mainz, Germany}
\affiliation{Ludwig-Maximilians-Universit{\"a}t M{\"u}nchen, M{\"u}nchen, Germany}
\affiliation{Fachbereich Physik, Bergische Universit{\"a}t Wuppertal, Wuppertal, Germany}
\affiliation{Panjab University, Chandigarh, India}
\affiliation{Delhi University, Delhi, India}
\affiliation{Tata Institute of Fundamental Research, Mumbai, India}
\affiliation{University College Dublin, Dublin, Ireland}
\affiliation{Korea Detector Laboratory, Korea University, Seoul, Korea}
\affiliation{CINVESTAV, Mexico City, Mexico}
\affiliation{Nikhef, Science Park, Amsterdam, the Netherlands}
\affiliation{Radboud University Nijmegen, Nijmegen, the Netherlands and Nikhef, Science Park, Amsterdam, the Netherlands}
\affiliation{Joint Institute for Nuclear Research, Dubna, Russia}
\affiliation{Institute for Theoretical and Experimental Physics, Moscow, Russia}
\affiliation{Moscow State University, Moscow, Russia}
\affiliation{Institute for High Energy Physics, Protvino, Russia}
\affiliation{Petersburg Nuclear Physics Institute, St. Petersburg, Russia}
\affiliation{Instituci\'{o} Catalana de Recerca i Estudis Avan\c{c}ats (ICREA) and Institut de F\'{i}sica d'Altes Energies (IFAE), Barcelona, Spain}
\affiliation{Stockholm University, Stockholm and Uppsala University, Uppsala, Sweden}
\affiliation{Lancaster University, Lancaster LA1 4YB, United Kingdom}
\affiliation{Imperial College London, London SW7 2AZ, United Kingdom}
\affiliation{The University of Manchester, Manchester M13 9PL, United Kingdom}
\affiliation{University of Arizona, Tucson, Arizona 85721, USA}
\affiliation{University of California Riverside, Riverside, California 92521, USA}
\affiliation{Florida State University, Tallahassee, Florida 32306, USA}
\affiliation{Fermi National Accelerator Laboratory, Batavia, Illinois 60510, USA}
\affiliation{University of Illinois at Chicago, Chicago, Illinois 60607, USA}
\affiliation{Northern Illinois University, DeKalb, Illinois 60115, USA}
\affiliation{Northwestern University, Evanston, Illinois 60208, USA}
\affiliation{Indiana University, Bloomington, Indiana 47405, USA}
\affiliation{Purdue University Calumet, Hammond, Indiana 46323, USA}
\affiliation{University of Notre Dame, Notre Dame, Indiana 46556, USA}
\affiliation{Iowa State University, Ames, Iowa 50011, USA}
\affiliation{University of Kansas, Lawrence, Kansas 66045, USA}
\affiliation{Kansas State University, Manhattan, Kansas 66506, USA}
\affiliation{Louisiana Tech University, Ruston, Louisiana 71272, USA}
\affiliation{Boston University, Boston, Massachusetts 02215, USA}
\affiliation{Northeastern University, Boston, Massachusetts 02115, USA}
\affiliation{University of Michigan, Ann Arbor, Michigan 48109, USA}
\affiliation{Michigan State University, East Lansing, Michigan 48824, USA}
\affiliation{University of Mississippi, University, Mississippi 38677, USA}
\affiliation{University of Nebraska, Lincoln, Nebraska 68588, USA}
\affiliation{Rutgers University, Piscataway, New Jersey 08855, USA}
\affiliation{Princeton University, Princeton, New Jersey 08544, USA}
\affiliation{State University of New York, Buffalo, New York 14260, USA}
\affiliation{Columbia University, New York, New York 10027, USA}
\affiliation{University of Rochester, Rochester, New York 14627, USA}
\affiliation{State University of New York, Stony Brook, New York 11794, USA}
\affiliation{Brookhaven National Laboratory, Upton, New York 11973, USA}
\affiliation{Langston University, Langston, Oklahoma 73050, USA}
\affiliation{University of Oklahoma, Norman, Oklahoma 73019, USA}
\affiliation{Oklahoma State University, Stillwater, Oklahoma 74078, USA}
\affiliation{Brown University, Providence, Rhode Island 02912, USA}
\affiliation{University of Texas, Arlington, Texas 76019, USA}
\affiliation{Southern Methodist University, Dallas, Texas 75275, USA}
\affiliation{Rice University, Houston, Texas 77005, USA}
\affiliation{University of Virginia, Charlottesville, Virginia 22901, USA}
\affiliation{University of Washington, Seattle, Washington 98195, USA}
\author{V.M.~Abazov} \affiliation{Joint Institute for Nuclear Research, Dubna, Russia}
\author{B.~Abbott} \affiliation{University of Oklahoma, Norman, Oklahoma 73019, USA}
\author{B.S.~Acharya} \affiliation{Tata Institute of Fundamental Research, Mumbai, India}
\author{M.~Adams} \affiliation{University of Illinois at Chicago, Chicago, Illinois 60607, USA}
\author{T.~Adams} \affiliation{Florida State University, Tallahassee, Florida 32306, USA}
\author{G.D.~Alexeev} \affiliation{Joint Institute for Nuclear Research, Dubna, Russia}
\author{G.~Alkhazov} \affiliation{Petersburg Nuclear Physics Institute, St. Petersburg, Russia}
\author{A.~Alton$^{a}$} \affiliation{University of Michigan, Ann Arbor, Michigan 48109, USA}
\author{G.~Alverson} \affiliation{Northeastern University, Boston, Massachusetts 02115, USA}
\author{G.A.~Alves} \affiliation{LAFEX, Centro Brasileiro de Pesquisas F{\'\i}sicas, Rio de Janeiro, Brazil}
% ...
%
%
\collaboration{The D0 Collaboration\footnote{with visitors from
$^{a}$Augustana College, Sioux Falls, SD, USA,
$^{b}$The University of Liverpool, Liverpool, UK,
$^{c}$UPIITA-IPN, Mexico City, Mexico,
$^{d}$SLAC, Menlo Park, CA, USA,
$^{e}$University College London, London, UK,
$^{f}$Centro de Investigacion en Computacion - IPN, Mexico City, Mexico,
$^{g}$ECFM, Universidad Autonoma de Sinaloa, Culiac\'an, Mexico,
and 
$^{h}$Universit{\"a}t Bern, Bern, Switzerland.
$^{\ddag}$Deceased.
}} \noaffiliation
\vskip 0.25cm
\date{October 20, 2011}

\begin{abstract}

We present a measurement of the relative branching fraction, $R_{f_{0}/\phi}$, of $B_{s}^{0} \rightarrow J/\psi f_{0}(980)$, 
with $f_{0}(980) \rightarrow \pi^{+}\pi^{-}$, to the process
$B_{s}^{0} \rightarrow J/\psi \phi$, with $\phi \rightarrow K^{+}K^{-}$.
The $J/\psi f_{0}(980)$ final state corresponds to a  
CP-odd eigenstate of $B_{s}^{0}$ that could be of interest in future studies of CP violation.  Using
8~$\rm{fb}^{-1}$ of data recorded with the D0 detector at the Fermilab Tevatron Collider, we find
$R_{f_{0}/\phi}$ = 0.275 $\pm$ 0.041\thinspace(stat) $\pm$ 0.061\thinspace(syst).

\end{abstract}


\maketitle


The CP-violating phase angle in $B_{s}^{0}$ mixing, $\phi_{s}^{J/\psi \phi}$,
has been measured \cite{d0phis,d0phis2,cdfphis} using  $B_{s}^{0} \rightarrow J/\psi \phi$ decays, and is statistically
consistent with that predicted by the standard model (SM) \cite{SMphi}. 
Ignoring possible ambiguities in its hadronic structure \cite{Fleischer}, the weak phase angle
$\phi_s^{c\bar{c} s\bar{s}}$ measured in $B_s \rightarrow J/\psi f_{0}(980)$ decay should be equal to the angle 
$\phi_s^{J/\psi \phi} = -2\beta_s^{SM} + \phi_s^{NP}$, where $\beta_s^{SM}$ is the SM angle in
the unitarity triangle for the $B^0_s$ system that is analogous to the well-known angle $\beta$ in the $B^0_d$ system, and $\phi_s^{NP}$ 
is any additional phase arising from
new physics in $B^0_s$ mixing. Measuring this phase angle through various decay modes could help reduce its uncertainty. 
In particular, the decay products of $B^0_s \rightarrow J/\psi \phi$ are in an indefinite CP state, requiring 
CP-even and CP-odd components to be extracted through a time-dependent angular analysis. 
In contrast, the decay products in $B^0_s \rightarrow J/\psi f_{0}(980)$ are in a CP-odd eigenstate, 
which can provide a more direct measurement of 
$\phi_s^{c\bar{c}s\bar{s}}$ relative
to $B^0_s \rightarrow J/\psi \phi$ \cite{cpodd}. The precision of a $\phi_s^{c\bar{c}s\bar{s}}$
measurement in the $f_{0}(980)$ channel is expected to be poorer than in the $\phi$ channel because of the smaller decay branching ratio 
for this process. However, a complementary method of analysis provides different systematic uncertainties, as well as an 
important cross check of the result from $B^0_s \rightarrow J/\psi \phi$.

In this Article, we present a measurement of the relative branching fraction ($R_{f_{0}/\phi}$) which is, based
on theoretical estimates, expected to be 
significant \cite{Stone,StoneZhang,Colangelo,Leitner,cleo}: 
\begin{equation}
R_{f_{0}/\phi} \equiv \frac { \Gamma (B_{s}^{0}\rightarrow J/\psi f_{0}(980),f_{0}(980)\rightarrow \pi^{+} \pi^{-}) } 
                  { \Gamma (B_{s}^{0}\rightarrow J/\psi \phi,\phi \rightarrow K^{+} K^{-}    )} \nonumber \end{equation}

\vspace{-0.15in}

\begin{equation}
\approx 0.20-0.40. 
\end{equation}
The LHCb Collaboration has reported \cite{LHCb} 
a first measurement of $R_{f_{0}/\phi}= 0.252^{+0.046}_{-0.032} \thinspace {\mathrm{(stat)}} ^{+0.027}_{-0.033} \thinspace {\mathrm{(syst)}}.$
The Belle Collaboration has measured the branching
fraction $\mathcal{B} (B_{s}^{0}\rightarrow J/\psi  f_{0}(980), 
 f_{0}(980) \rightarrow \pi^{+} \pi^{-})$ = 
$[1.16^{+0.31}_{-0.19}\thinspace({\mathrm{stat}})^{+0.15}_{-0.17}\thinspace({\mathrm{syst}})^{+0.26}_{-0.18}(N_{B_{s}^{(*)}\bar{B}_{s}^{(*)}})] 
\times 10^{-4}$ \cite{Bellemeasure}, where $N_{B_{s}^{(*)}\bar{B}_{s}^{(*)}}$ is the number of $B_{s}^{(*)}\bar{B}_{s}^{(*)}$ pairs in the sample.
The CDF Collaboration has also measured the relative branching fraction and finds $R_{f_{0}/\phi}=0.257$ $\pm$ 0.020\thinspace(stat) $\pm$ 0.014\thinspace(syst) 
\cite{CDFf0}.
We report a new measurement of the relative branching fraction using data collected with the D0 detector at the Fermilab
Tevatron Collider.


To determine an absolute branching fraction requires an excellent understanding of efficiencies, other related
branching fractions, cross sections, 
and integrated luminosity. However, by measuring a relative 
branching fraction, terms common to both the $B_{s}^{0}\rightarrow J/\psi f_{0}(980)$ and the $B_{s}^{0}\rightarrow J/\psi \phi$ decays cancel, giving:
\begin{eqnarray}
R_{f_{0}/\phi}  & = & \frac{N_{ B_{s}^{0} \rightarrow J/\psi f_{0}(980)}} {N_{B_{s}^{0} \rightarrow J/\psi \phi}} 
         \cdot 
        \frac {\varepsilon_{\mathrm{reco}}^{B_{s}^{0} \rightarrow J/\psi \phi}}{\varepsilon_{\mathrm{reco}}^{B_{s}^{0} \rightarrow J/\psi f_{0}(980)} }\;.
\end{eqnarray}
Hence, just the yields 
$N_{ B_{s}^{0} \rightarrow J/\psi f_{0}(980)}$ and 
$N_{B_{s}^{0} \rightarrow J/\psi \phi}$  and their detection efficiencies, 
$\varepsilon_{\mathrm{reco}}^{B_{s}^{0} \rightarrow J/\psi \phi}$ 
and $\varepsilon_{\mathrm{reco}}^{B_{s}^{0} \rightarrow J/\psi f_{0}(980)}$,
are needed to measure $R_{f_{0}/\phi}$.


The D0 detector is described in Ref.\cite{d0det}, and
only those components that directly affect this measurement are
discussed below.  The tracking system consists of a silicon microstrip tracker
(SMT) and a central fiber tracker (CFT), both located within a 1.9~T
superconducting solenoid magnet. The SMT has approximately 
800,000 individual strips, with typical pitch of $50-80$ $\mu$m, and 
a design optimized for tracking and vertexing capability within the pseudorapidity range 
$|\eta|<3$ \cite{eta}.
The system has a six-barrel longitudinal structure, each barrel having 
four layers arranged axially around the beam pipe, interspersed 
with 16 radial disks. The CFT has eight thin coaxial barrels, each 
supporting two doublets of overlapping scintillating fibers of 
0.835~mm diameter.  One doublet is parallel to the collision 
axis, and the others alternate by $\pm 3^{\circ}$ relative to that axis. 
The muon system resides beyond a calorimeter that surrounds the inner tracking detectors, 
and consists of one layer of tracking detectors and scintillation trigger counters 
before 1.8~T toroids, followed by two similar layers after
the toroids. 


Approximately 8 fb$^{-1}$ of integrated luminosity 
is used in this analysis.
The data are
divided into four time periods, corresponding to different detector configurations and instantaneous
luminosities,
called Run IIa (1.4 fb$^{-1}$), Run IIb1 (1.4 fb$^{-1}$), Run IIb2 (3.3 fb$^{-1}$), and Run IIb3 (2.1 fb$^{-1}$).

We search for $B_{s}^{0} \rightarrow J/\psi f_{0}(980)$ candidates using the decay mode
$J/\psi \rightarrow \mu^{+} \mu^{-}$.
Events are collected using a mixture of single and
dimuon triggers which have a similar trigger efficiency for both
$B_{s}^{0} \rightarrow J/\psi f_{0}(980)$ and $B_{s}^{0} \rightarrow J/\psi \phi$.
Muon candidates must have transverse momentum $p_{T}$ $>$ 1.5 GeV and 
be detected in the muon chambers within the toroidal magnet. 
In addition, each muon track must
be associated with a track reconstructed by the CFT, and
have at least one SMT hit.
The $J/\psi$ candidates are formed from two oppositely charged muon
candidates emanating from a common vertex, and have at least one of the
muon candidates detected outside the toroidal magnet.

All reconstructed tracks not associated with muons forming a $J/\psi$ candidate are considered 
in the reconstruction of  $f_{0}(980)$ and $\phi$ candidates.  
Since the D0 detector has limited ability to separate kaons from pions, tracks 
are assigned the pion mass when searching for
$B_{s}^{0} \rightarrow J/\psi f_{0}(980)$ and the kaon mass when searching for  $B_{s}^{0} \rightarrow J/\psi \phi$.
Charged tracks are required to have at least two CFT hits, at least
two SMT hits, a total of at least eight SMT and CFT hits, and a minimum $p_{T}$ of 300 MeV.
Any two oppositely charged tracks that have one track with transverse momentum $p_{T}$ $>$ 1.4 GeV, 
an invariant mass 
0.7 GeV $<$ $M_{\pi^{+} \pi^{-}}$ $<$ 1.2 GeV or 1.0 GeV $<$ $M_{K^{+} K^{-}}$ $<$ 1.05 GeV, 
and are consistent with originating from a common vertex,
are considered as $f_{0}(980)$ and $\phi$ candidates, respectively. 
The $\mu^{+} \mu^{-} \pi^{+}\pi^{-}$ ($\mu^{+} \mu^{-} K^{+}K^{-}$) candidates are required to form a common 
vertex and have an invariant mass between 5.0 and 5.8~GeV.
The invariant mass requirements 
on $M_{\pi^{+} \pi^{-}}$ and $M_{K^{+} K^{-}}$ prevent the two tracks to be considered as candidates for 
both $f_{0}(980)$ and $\phi$ interpretations.

The final data sample is formed by applying the following additional requirements to further
reduce backgrounds.
The $f_{0}(980)$  and $\phi$  candidates must have $p_{T}$ $>$ 1.6 GeV with
 0.91 GeV $<$ $M_{\pi^{+}\pi^{-}}$ $<$ 1.05 GeV and 1.01 GeV $<$ $M_{K^+K^-}$ $<$ 1.03 GeV.
The $B_{s}^{0}$ candidates are required to 
have $p_{T}$ $>$ 5 GeV,
 2.9 GeV $<$ $M_{\mu^+\mu^-}$ $<$ 3.2~GeV,
and have a proper decay length with a significance of greater than 5 standard deviations (sd).

The proper decay length, defined as $L_{xy}\cdot(M_{B_{s}^{0}}/p_{T})$, where $p_{T}$ is the 
transverse momentum of the $B_{s}^{0}$, 
$M_{B_{s}^{0}}$ is the PDG value of the mass of the $B_{s}^{0}$ \cite{PDG}, and $L_{xy}$ 
\cite{lxy} is the transverse 
distance between the primary $p\bar{p}$ interaction vertex 
and the four-track vertex of the $B_{s}^{0}$ candidate, is calculated for candidate primary vertices that use
the transverse beamspot as a constraint.
If there is more than one such vertex in an event, the primary vertex nearest in the transverse plane to the $J/\psi$ candidate is 
chosen for this analysis.

A final selection is based on two Boosted Decision Tree \cite{BDT1,BDT2} (BDT) discriminants.
We use the Monte Carlo (MC) {\sc pythia} program \cite{pythia} to generate $B_{s}^{0}$ events and the {\sc evtgen} program \cite{Evtgen} to
simulate their decay.
MC signal and background samples are used to train a BDT  and to form discriminant output values for each event.
The expected background is primarily due to two sources: prompt background that is defined as directly produced $J/\psi$ mesons 
accompanied by tracks from hadronization, and non-prompt, or inclusive $B \rightarrow J/\psi +X$ decays where the  $J/\psi$ meson arises
from a $b$-hadron decay accompanied by tracks from hadronization.
Two MC background samples are therefore generated with {\sc pythia}:
a sample of directly produced $J/\psi$ prompt events and an inclusive sample of $B_{s}^{0}$ for all decay processes
$B^0_{s}\rightarrow J/\psi + X$.  A MC signal sample of $J/\psi f_{0}(980)$ events is used 
to train both BDTs.
Thirty input variables are used in the BDT, including the momenta of final-state objects, 
vertex-quality requirements, $B_{s}^{0}$ isolation, and decay angles.
Six BDT isolation variables are used in this BDT, representing different choices for the size
and which tracks are included in the cone of isolation. 
The BDT selections for both prompt and inclusive training are defined with a requirement on the BDT output value
which provides large $S$/$\sqrt{B}$,  while keeping the signal yields high, where $S$ and $B$ are the number of
signal and background events.

\begin{figure}
\includegraphics[scale=0.4]{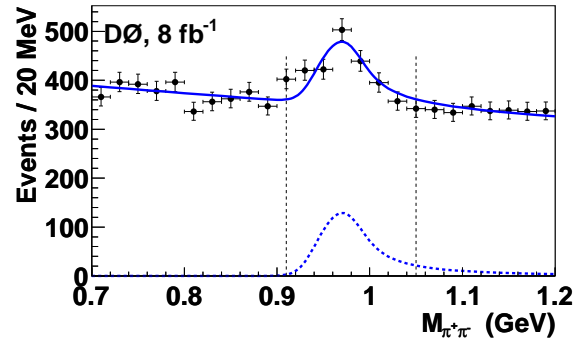}
\caption{\label{fig:f0} The invariant mass distribution of 
$f_0(980)$ candidates when the $J/\psi \pi^{+}\pi^{-}$ invariant 
mass is within $\pm$2 sd of the fitted mean $B^0_{s}$ mass.  The solid line represents the fit to all the data,
and the
dashed line the fitted $f_0(980)$ signal (see text).
The vertical dashed lines indicate the region 0.91 GeV $<$ $M_{\pi^+\pi^-}$ $<$ 1.05 GeV.}
\end{figure}

\begin{figure}
\includegraphics[scale=0.4]{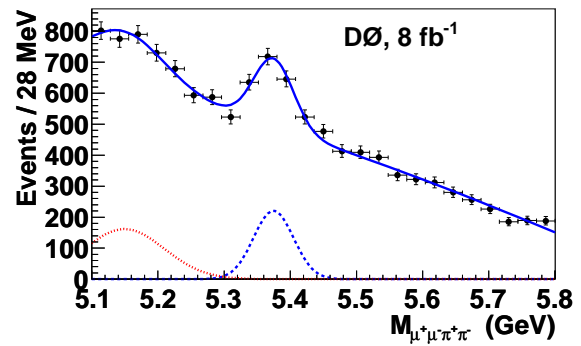}
\caption{\label{fig:signal} The invariant  mass distribution 
of $B^0_s$ candidates when the
 $\pi^{+}\pi^{-}$ invariant mass is consistent with
that of a $f_0(980)$ meson, i.e., 0.91 $<$ $M_{\pi^+ \pi^-}$ $<$ 1.05 GeV.
The solid line is the fit to all the data and the
dashed line the fitted $B^0_s$ signal.  The dotted line is a Gaussian function used to describe
partially reconstructed $B$ decays. 
(see text).}
\end{figure}

The invariant masses of $f_{0}(980)$ and
$B_{s} \rightarrow J/\psi f_{0}(980)$ candidates, following
BDT selections are shown in
Figs.~\ref{fig:f0} and \ref{fig:signal}, respectively.
An unbinned likelihood fit is used to determine the yield of signal in each sample.
The $f_{0}(980)$ has a large width \cite{PDG} and a mass just below the $KK$ threshold.  
This affects the line shape, which is not
a simple Breit-Wigner form, particularly at large mass values.   
The $\pi^{+}\pi^{-}$ mass distribution is therefore fitted using the functional form 
of Ref. \cite{Flatte}, which takes account of the opening of the $KK$ threshold, and is
convoluted with a Gaussian resolution function with a sd of 15 MeV.
The line shape determined by fitting the $f_{0}(980)$ in MC, using a second-degree polynomial for the background 
is also used to fit the data. 
Candidates for $B_{s}^{0}\rightarrow J/\psi f_{0}(980)$ are defined by the $\pi^{+}\pi^{-}$ invariant mass
window 0.91 $<$ $M_{\pi^+\pi^-}$ $<$ 1.05 GeV.
The  $B_{s}^{0} \rightarrow J/\psi f_{0}(980)$ mass distribution is 
fitted to a Gaussian signal, with a background function consisting
of a second-degree polynomial and a Gaussian at lower invariant mass to take into account partially 
reconstructed $B$ decays.  The unbinned likelihood fit is used to determine the contribution to signal in each sample.  
The $J/\psi f_{0}(980)$ mass distribution shown in Fig.~\ref{fig:signal} yields a fitted
$ B_{s}^{0}$ mass of  5.3748 $\pm$ 0.0036 GeV and  590 $\pm$ 84 $B_{s}^{0}$ events, where the
uncertainties reflect just the statistical uncertainties on the fit.

\begin{figure}
\includegraphics[scale=0.4]{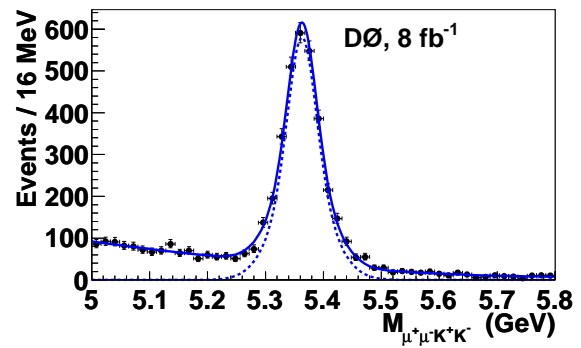} 
\caption{\label{fig:phi} The invariant $J/\psi K^{+}K^{-}$ mass distribution when the
 $K^{+}K^{-}$ invariant mass is consistent with a $\phi$ meson, i.e.,  1.01 $<$ $M_{K^+ K^-}$ $<$ 1.03 GeV.
 The solid line is the fit to all the data and the
dashed line is the part fitted to $B^0_s$ signal.
(see text).}

\end{figure}

Using identical event selections, except for the $f_{0}(980)$ mass
requirement, 
a clear $J/\psi \phi$ peak is found, as shown in Fig.~\ref{fig:phi}. 
The $\mu^+\mu^-K^+K^-$ mass distribution is fitted for a  $B_{s}^{0} \rightarrow J/\psi \phi$ signal
using a double Gaussian function with a second-order polynomial for background.
An unbinned likelihood fit to the $J/\psi \phi$ distribution shown in Fig.~\ref{fig:phi} yields a 
$ B_{s}^{0}$ mass of  5.3631 $\pm$ 0.0008 GeV and 2929 $\pm$ 62 $B_{s}^{0}$ events, where again the
uncertainties are statistical only.

MC signal samples are used to determine the efficiencies of reconstructing the two $B_{s}^{0}$ decay modes.
To take account of changes in the instantaneous
luminosity, the MC samples are overlaid with data events from random beam crossings collected during each run period.
In the generation of both the $B^0_s \rightarrow J/\psi \phi$ and the $B^0_s \rightarrow J/\psi f_{0}(980)$ MC signals, 
a preselection requirement of $p_{T}$
$>$ 0.4 GeV is imposed on both kaons and pions from the $\phi$ and $f_{0}(980)$.  Since the $p_{T}$ distributions
of pions and kaons differ, the preselection efficiencies are determined from two additional MC sets of events generated without
$p_{T}$ cutoffs.

Reconstruction efficiencies depend on the data-taking period (Run IIa -- IIb3)
as instantaneous luminosity, aging of the detector, and changes to the reconstruction 
algorithms affect detector performance. The reconstruction efficiencies are therefore measured
separately for each running period.
The instantaneous luminosities for data taken during 
Run IIb3 are similar to those of Run IIb2, and the reconstruction efficiencies for Run IIb2 are therefore also used for Run IIb3 data.
Although the absolute reconstruction efficiencies depend on the running period, 
the relative reconstruction efficiencies given in Table \ref{tab:ranges} are stable.  The differences in relative reconstruction efficiency
are used to estimate a systematic uncertainty on $R_{f_{0}/\phi}$.
The mean relative reconstruction efficiency is 1.20 $\pm$ 0.04, where the uncertainty is from
statistics in the MC.

The $B_{s}^{0} \rightarrow J/\psi \phi$ time development reflects 
a mix of two exponential functions with relative slope
values driven by the difference in decay widths of
the two mass eigenstates ($\Delta\Gamma_s$).  The relative efficiency of any cutoff on proper decay length for the
two states depends on $\Delta\Gamma_s$ and the lifetime of the CP-odd
eigenstate.
The MC samples used to determine the relative efficiency use
$\Delta \Gamma_s = 0$ and the PDG value of
$\Gamma_{s}$ \cite{PDG}.   For this
$\Delta \Gamma_s$, and assuming no CP violation, the effect on the relative efficiency of $f_{0}(980)/\phi$ is
found to be small ($\approx$2.5\%) and well within systematic uncertainties, and therefore no correction is applied.


\begin{table}
\caption{\label{tab:ranges} Relative reconstruction efficiencies for different running periods.}
\begin{ruledtabular}
\begin{tabular}{lc}

\hline
Run period & $ \varepsilon_{reco}^{B_{s}^{0} \rightarrow J/\psi \phi} / \varepsilon_{reco}^{B_{s}^{0} \rightarrow J/\psi f_{0}(980)} $ \\
\hline
Run IIa  & 1.19 $\pm$ 0.03 \\
Run IIb1 & 1.29 $\pm$ 0.04 \\
Run IIb2+IIb3 & 1.20 $\pm$ 0.05 \\
\end{tabular}
\end{ruledtabular}
\end{table}



The branching fraction of $B^{0}_{s} \rightarrow J/\psi f_{0}(980)$ is measured relative
to $B^{0}_{s} \rightarrow J/\psi \phi$, so any backgrounds that peak under the $B^{0}_{s} \rightarrow J/\psi \phi$ 
mass distribution will affect the measurement of $R_{f_{0}/\phi}$. 
Possible $\mathcal{S}$-wave contributions can arise from the $f_{0}$ or from non-resonant
$K^{+}K^{-}$ production, but these contributions provide only slowly varying contributions under the $\phi$
mass peak.  The excess for larger $M_{K^{+}K^{-}}$
is extrapolated under the $\phi$ mass, giving a possible $\mathcal{S}$-wave contribution of (12 $\pm$ 3)\% \cite{swave} of the 
total $B_{s}^{0} \rightarrow J/\psi \phi$ yield.  The relative branching ratio is therefore scaled up by a factor of 1/0.88 to
account for an $\mathcal{S}$-wave contribution to $B_{s}^{0} \rightarrow J/\psi \phi$.



One possible background that can affect the observed $B_{s}^{0} \rightarrow J/\psi f_{0}(980)$ yield 
is the three-body decay $B_{s}^{0} \rightarrow J/\psi \pi^{+} \pi^{-}$.  
This background is studied by measuring the $B_{s}^{0}$ yield for $\pi^{+} \pi^{-}$ invariant masses less
than the $f_{0}(980)$ mass.  The $\pi^{+} \pi^{-}$ mass distribution from non-resonant $B_{s}^{0} \rightarrow J/\psi 
\pi^{+} \pi^{-}$ is broad, and measuring the $B_{s}^{0}$ yield
for a sideband in $M_{\pi^{+} \pi^{-}}$ therefore provides an extrapolation of the non-resonant $\pi^{+} \pi^{-}$ background
to the $f_{0}(980)$ signal region.
In defining a $\pi^{+} \pi^{-}$ mass window to study this background, it is important to avoid regions where
other known resonances, e.g., $B_{s}^{0} \rightarrow J/\psi K^{*}, K^{*} \rightarrow K\pi$ (with the kaon
assigned the pion mass) can contribute.  
The  $\pi^{+} \pi^{-}$ mass window of 0.8--0.9 GeV
is chosen because this mass range has no overlap with $B_{s}^{0} \rightarrow J/\psi K^{*}$ events.  

The mass distribution of $\mu^{+} \mu^{-} \pi^{+}\pi^{-}$, for 0.8 $<$ $M_{\pi^{+}\pi^{-}}$ $<$ 0.9 GeV,
is fitted with a floating contribution from non-resonant $J/\psi \pi^+ \pi^-$ decays.
The mean and the width of the $B^0_s$ peak are constrained to the
values obtained from the corresponding fit in the $f_0(980)$ signal region.
The fit yields 42 $\pm$ 49 events, indicating no evidence of 
$B_{s}^{0} \rightarrow J/\psi \pi^{+} \pi^{-}$ non-resonant background, and consequently no
such correction is used in this analysis.




To check that the results of the analysis do not depend on the specific
choice of the selection critera, each cut is changed around its nominal
value, and it is observed that $R_{f_{0}/\phi}$ does not depend significantly on the exact choice of
selections. 

The large backgrounds arising from particle combinatorics and from partially reconstructed $B$ decays provide 
significant distortion and uncertainties in the distributions of background. We study this using same-charge pions and 
the mass distribution from
$\mu^{+} \mu^{-} \pi^{\pm} \pi^{\pm}$ events.  However, we find that the $\mu^{+} \mu^{-} \pi^{\pm} \pi^{\pm}$ 
distribution does not describe the
measured background in our signal sample and we therefore do not use it to help constrain the distribution
of the background.  Instead, 
different parameterizations are used (third-degree polynomial and
an exponential) to
describe the background, and different mass regions over which the fit is performed are used 
to determine the signal yield variation.
A large variation in the number 
of signal events for $B_{s}^{0}\rightarrow J/\psi f_{0}(980)$ is found for different parameterizations of
the background, 
indicating that modeling of the background has substantial ambiguity.
The choice of background
parametrization comprises the largest contribution to the total 
systematic uncertainty
on $R_{f_{0}/\phi}$.

A similar study of fitting choices is performed on the $B_{s}^{0}\rightarrow J/\psi \phi$ sample. 
However, since these backgrounds are much smaller and
easier to describe, the measured event yields change by less than 1\%.  
The presence of a $B^0 \rightarrow J/\psi \pi^+ \pi^-$ contribution is checked by including this channel
in the fit, yielding a fit consistent with no events.

The MC distributions of the kinematic variables do not describe the data perfectly in all variables. To study this
effect on the training of the BDT, the MC distributions for signal are weighted to match 
the $B_{s}^{0}\rightarrow J/\psi \phi$ data.
Only the $B_{s}^{0}\rightarrow J/\psi \phi$ events are used for this purpose because 
in the  $B_{s}^{0}\rightarrow J/\psi f_{0}(980)$ channel there is much 
background and a far smaller signal fraction.

Using the Run IIb2 data and Run IIb2 MC, we find that the relative efficiency for event reconstruction changes from
1.20 $\pm$ 0.05 without reweighting to 2.00 $\pm$ 0.07 after weighting.  Although this corresponds to a large difference in
relative efficiency, the relative yields also change, thereby changing  $R_{f_{0}/\phi}$ by just 17.8\%.
Half of the difference between the nominal result and the reweighted BDT result is taken as a systematic uncertainty
on $R_{f_{0}/\phi}$.
A 4.0\% systematic uncertainty is assigned for the observed dependence of $R_{f_{0}/\phi}$  
on the size of the $f_0(980)$ mass window.
Table \ref{tab:unc} summarizes the values of the systematic uncertainties on $R_{f_{0}/\phi}$.

\begin{table}
\caption{\label{tab:unc} Systematic uncertainties in the branching fraction ratio, $R_{f_{0}/\phi}$.}
\begin{ruledtabular}
\begin{tabular}{lc}

\hline
Source & Uncertainty \\
\hline
Fitting & 17.3\% \\
MC efficiency & 9.2\% \\
Modeling variables in BDT & 8.9\% \\ 
$f_0(980)$ mass window & 4.0\% \\
 $\mathcal{S}$-wave contribution & 3.5\% \\
\hline
Total & 22.2\% \\
\end{tabular}
\end{ruledtabular}
\end{table}


Based on  8~fb$^{-1}$ of data, D0 has extracted a measurement of the relative branching fraction  $R_{f_{0}/\phi}$ 
of Eq. 1. 

$$ R_{f_{0}/\phi} =  0.275 \pm 0.041\thinspace(\rm{stat}) \pm 0.061\thinspace (\rm{syst}). $$

This agrees with theoretical expectations and with previous measurements of the ratio of widths.



%
We thank the staffs at Fermilab and collaborating institutions,
and acknowledge support from the
DOE and NSF (USA);
CEA and CNRS/IN2P3 (France);
FASI, Rosatom and RFBR (Russia);
CNPq, FAPERJ, FAPESP and FUNDUNESP (Brazil);
DAE and DST (India);
Colciencias (Colombia);
CONACyT (Mexico);
KRF and KOSEF (Korea);
CONICET and UBACyT (Argentina);
FOM (The Netherlands);
STFC and the Royal Society (United Kingdom);
MSMT and GACR (Czech Republic);
CRC Program and NSERC (Canada);
BMBF and DFG (Germany);
SFI (Ireland);
The Swedish Research Council (Sweden);
and
CAS and CNSF (China).
%


\begin{thebibliography}{99}
\bibitem{eta} Pseudorapidity is defined as $\eta = -\ln[\tan(\theta/2)]$, where $\theta$ is
the polar angle relative to the proton-beam direction.
\end{thebibliography}
\end{document}